\documentclass[twocolumn,12pt,cites,graphicx]{article}
\usepackage{epsfig}
\usepackage{color}

\title{Lattice Boltzmann simulations of liquid crystalline fluids: 
active gels and blue phases}

\author{M. E. Cates$^1$, O. Henrich$^1$, D. Marenduzzo$^1$, K. Stratford$^2$ \\
$^1$SUPA, School of Physics and Astronomy, and $^2$EPCC, \\ University of Edinburgh, JCMB Kings Buildings,\\
Mayfield Road, Edinburgh EH9 3JZ, Scotland}

\begin{document}

\maketitle

\noindent
Lattice Boltzmann simulations have become a method of choice
to solve the hydrodynamic equations of motion of a number of
complex fluids. Here we review some recent applications of lattice Boltzmann
to study the hydrodynamics of liquid crystalline materials. In particular, we
focus on the study of (a) the exotic blue phases of cholesteric liquid crystals, and (b) active gels -- a model system for actin plus myosin solutions or
bacterial suspensions. In both cases lattice Boltzmann studies have
proved useful to provide new insights into these complex materials.


\section{Introduction}

In recent years, the lattice Boltzmann (LB) algorithm~\cite{Succi} 
has emerged as a powerful method to study fluid dynamics. Due to its 
conceptual simplicity and codability (particularly on parallel computers), LB provides an attractive alternative 
to other methods such as finite elements algorithms. In the 
last couple of decades, in particular,
the LB method has increasingly been applied to
the hydrodynamics of complex fluids, such as binary fluids, colloidal 
suspensions and liquid crystals~\cite{bijel_review,Swift96,Gonnella97,Denniston04,Marenduzzo07a,Cates08,Stratford}. 
  
In applying LB to complex fluids, one often aims at solving
two coupled sets of partial differential 
equations. One set describes the evolution of the
order parameter (e.g., composition for binary mixtures, 
or orientational order in liquid crystals), whereas the other set describes conservation of mass and momentum (via the continuity and Navier-Stokes equations for the velocity field).
In liquid crystals, which are the focus of this article, one typically considers the 
Beris-Edwards model~\cite{Beris}, which is defined starting from a 
free energy expressed in terms of a tensorial order parameter, ${\bf Q}$, whose largest 
eigenvalue describes locally the strength of local molecular alignment (nematic order), and whose corresponding eigenvector defines the director $\mathbf{n}$ along which this alignment prevails. (See the Appendix for the precise form of the free 
energy adopted.) Note that simpler descriptions, in which the magnitude of the ordering is assumed constant and only the director varies, are unsuitable for describing blue phases. This is because such phases contain defect lines at which the ordering drops locally \cite{Wright89}.

The equation of motion for $\mathbf{Q}$ is then~\cite{Beris}:
\begin{equation}\label{Q_motion}
D_t \mathbf{Q} = \Gamma \mathbf{H}.
\end{equation}
The left member of equation \ref{Q_motion} is a ``material derivative'' describing the time evolution of the order parameter advected with the velocity $\mathbf{u}$ of a fluid element.
For fluids of rod-like particles such as liquid crystals, 
flow gradients may lead to local rotations and these are also
taken into account by the material derivative.
In Eq. \ref{Q_motion}, $\Gamma$ is a collective rotational diffusion constant, which sets the time scale for the relaxation of orientational order 
(usually in the millisecond range for small-molecule liquid crystals),
and $\mathbf{H}$ is the ``molecular field'', which provides the force for this 
relaxation motion. The molecular field involves the derivative of the 
free energy with respect to the order parameter (see the Appendix for the specific form of $\mathbf{H}$).

As stated previously, the fluid velocity obeys the continuity equation, 
$\partial_t\rho = -\nabla.\mathbf{u} \equiv \partial_\alpha u_\alpha$ where $\rho$ is the fluid density, which in practical cases this can be taken as constant (so that the fluid is incompressible). However, a feature of LB is that slight fluid compressibility is maintained in the algorithm; this makes all the dynamics fully local (the sound-speed is finite) rather than having to solve each timestep for a pressure field that responds instantaneously to distant events. (Locality is important to efficient parallelization strategies, and maintains near-linear scaling of the computational cost with the size of the system under investigation \cite{Kevinscaling}.) 

The fluid velocity also obeys the Navier-Stokes 
equation, which for effectively incompressible fluids reads
\begin{eqnarray}\label{Navier-Stokes}
  \rho\left[\partial_t + u_\beta\partial_\beta\right] u_\alpha & = &
  \partial_\beta \Pi_{\alpha\beta} \\ \nonumber
  & + & \eta\partial_{\beta}\left(\partial_\alpha u_\beta+
\partial_\beta u_\alpha\right),
\end{eqnarray}
where $\eta$ is the viscosity, Cartesian components are denoted by Greek indices, 
and $\Pi_{\alpha\beta}$ is a thermodynamic stress tensor. This tensor, like $\mathbf{H}$, is found by differentiation of the free energy, and its divergence $\partial_\beta\Pi_{\alpha\beta}$ represents an effective body force, acting on the fluid, arising from the response to deformation of the order parameter field. (Appendix A gives the explicit form.) In the case of active fluids, the forces do not solely stem from a free energy but include terms arising from the dissipative conversion of chemical energy into motion. This creates an additional contribution to $\Pi_{\alpha\beta}$.   

Eqs. \ref{Q_motion} and \ref{Navier-Stokes} are very complex to solve due to their
inherent nonlinearities, and only limited progress is possible with
analytical techniques. In contrast, LB offers an ideal method to solve numerically these equations, allowing their full dynamics to be addressed not only in one dimension but also in 2D and 3D flows. Such studies can not only test approximate analytic solutions where these exist, but also lead to improved insight into the nonlinear physics contained in the underlying models.

For purely Newtonian fluids, the LB algorithm proceeds by the introduction of ``mesoscopic''
velocity distribution functions $f(\mathbf{c}_i;\mathbf{x})$, proportional to the density of (notional) fluid particles sitting at a lattice node at $\mathbf{x}$ having a certain velocity $\mathbf{c}_i$ chosen from a discrete set~\cite{Succi}. (These discrete velocities correspond to moving by one lattice site in a timestep.)
The distribution functions evolve 
according to an appropriate local dynamics, which recovers the Navier-Stokes 
and continuity equations upon coarse graining. (The density $\rho$ then equates to $\sum_i f(\mathbf{c}_i)$ and the average fluid velocity $\mathbf{u}$ becomes the first
moment, $\sum_i f(\mathbf{c}_i)\mathbf{c}_i$, of the distribution.) 

Historically, the first LB approach to
study complex fluids~\cite{Swift96,Denniston04} 
consisted of one extra set of
distribution functions for each new order parameter component entering
the equations of motion. 
More recently, however, it has become apparent that this ``full LB''
approach had the drawback of requiring a large memory to store all the 
distribution functions for the whole lattice, on top of being rather
cumbersome from a theoretical viewpoint. At the same time,
the new generation of LB studies for complex fluids
has shown the need to address ever larger systems,
to make best use of the potentially very good scalability of parallel LB codes.

As a result, new hybrid algorithms have been coded and deployed, for both
binary fluids and liquid crystals \cite{Tiribocchi09,Marenduzzo07b}, where 
the LB algorithm is used to solve the Navier-Stokes equation 
(Eq. \ref{Navier-Stokes}), and is coupled to a standard
finite-difference solver for the order parameter dynamics
(Eq. \ref{Q_motion}). At each timestep, the fluid velocity found by LB is used to calculate the advective derivative in Eq. \ref{Q_motion}, while the order parameter found from that equation is used to compute the forcing term, $\partial_\beta\Pi_{\alpha\beta}$, in Eq. \ref{Navier-Stokes}. By this division of labour, LB is \emph{only} used to handle the momentum and mass transport -- the problem for which it was originally devised. 

In what follows, we review recent applications of full and hybrid LB algorithms 
to the study of active liquid crystalline fluids and the blue phases of
cholesteric liquid crystals. In both cases, as we shall see, numerical
simulations have proved extremely helpful in providing a link between 
theoretical predictions and experimental observations. 

While in this work we focus on a ``hydrodynamic'' description
of liquid crystalline fluids via continuum models, it is important to
note that there is a variety of other coarse grained models and methods to
study liquid crystals. Most relevant to our topics here, liquid crystalline
molecules may be individually modelled via e.g. soft spherocylinders, 
interacting with Gay-Berne potentials (see e.g. \cite{PCCP09,Wilson08} 
for a series of recent examples) and a related approach has also managed to
stabilise blue phases~\cite{Memmer00}, although the length scale accessible
with this more detailed approach is typically significantly smaller than the
ones which can be studied with an LB continuum description.

\section{Active fluid simulations}

Active fluids have become a highly topical research area at the interface of soft matter and biological physics.
Generally speaking, an active particle ``absorbs energy from its surroundings or from an internal fuel 
tank and dissipates it in the process of carrying out internal movements''. (The quote is from a well-known paper of the Bangalore group, which helped pioneer the use of continuum models of active matter~\cite{Hatwalne04}.) 
This definition applies to bacteria, swimming algae and other microorganisms,
as well as to living cells or cell extracts. However active materials may also be non-biological, and a synthetic example is a shaken granular fluid~\cite{Narayan07}.
As the definition clearly implies, active materials may remain far from
thermodynamic equilibrium even in steady state, due to their continuous energy intake, and
this renders their properties of particular interest to physicists.

We consider active fluids comprising a concentrated suspension of such active particles.
Paradigmatic examples of active fluids are 
bacterial suspensions and solutions of cytoskeletal gel components (actin fibers
or microtubules) with molecular motors (myosin or kinesin). 
The term ``active gels'' is also widely used for these materials; but while all are non-Newtonian, not all of them are strongly viscoelastic.
Experimental studies with active fluids have uncovered a wide
range of intriguing and non-trivial physical properties. For instance,
microscopy studies of droplets of {\it Salmonella}~\cite{Harshey} and
of {\it B. subtilis}~\cite{Goldstein,Aranson,Cisneros07} 
reveal that when the bacteria are
concentrated enough (more than about 20-30\% in volume fraction),
long range correlations arise, creating eye-catching
patterns of flow involving long-lived vortices (see Fig. 1). These resemble the turbulent flow of fluids at high Reynold numbers, although remarkably in this case, the Reynolds number
is very small -- effectively zero~\cite{Bray}. (In this respect, the resulting ``bacterial turbulence'' resembles
elastic turbulence in polymer solutions, which is attained past 
a critical value of the Wiessenberg number~\cite{Morozov,Larson}.)

The equations of motion of an active liquid crystalline fluid have
been written down, either on the basis of symmetry~\cite{Hatwalne04,Kruse04}, 
or via a coarse graining of an underlying microscopic model of 
stiff cytoskeletal gels and molecular motors~\cite{Liverpool}. 
There are three main differences between these equations and 
the equations of motion of a 
passive liquid crystal. The first one is the presence of an active term in the 
stress tensor, whose divergence acts as a force in the Navier-Stokes equation (Eq. \ref{Navier-Stokes}). 
This extra active term has been first shown in \cite{Simha02} to be 
proportional to an activity constant, $\zeta$, which to lower order is linear
in the energy uptake of the fluid (e.g., via ATP hydrolysis), times
the local order parameter $\mathbf{Q}$. 
A second term arises from activity also, but the form of this can be absorbed into the free energy (see Appendix) and we set this to zero.
A third and final difference can arise in cases where an oriented swarm of swimming particles have a collective mean velocity relative to the surrounding fluid, causing a `self-advection' effect. Such materials are called `polar' and are distinct from the `apolar' case which describes either a swarm in which equal number of particles swim forward and backward along the director axis, or particles which are non-motile but nonetheless exert active forces on the fluid. (Motile and non-motile active particles are sometimes called ``movers'' and ``shakers'' respectively.)
More details on the exact forms of these active terms, and of the equations of motion used to describe polar systems, are given in the Appendix.

The sign of $\zeta$ -- hence of the active apolar contribution to the stress
tensor -- is of vital importance for the hydrodynamics and the 
rheology of active fluids~\cite{Marenduzzo07a,Cates08,Hatwalne04}.
A positive $\zeta$ corresponds to a suspension of extensile active particle,
or ``pushers''~\cite{Ishikawa08}, which exert forces along the molecular axis away from the centre of mass and
towards the surrounding fluid. A negative $\zeta$ corresponds to
a fluid of contractile active particles, or ``pullers''~\cite{Ishikawa08},
for which the force dipoles is exerted axially towards the centre of mass.
Examples of contractile fluids are suspensions of the biflagellated alga
{\it Chlamydomonas} and actomyosin gels (or more general suspensions
of non-permanently cross-linked cytoskeletal gels and molecular motors).
On the other hand, the majority of bacteria are thought to be 
extensile~\cite{Ishikawa08}. However, we are currently lacking 
quantitative experiments which measure the velocity field around active particles, which could lead to estimates of the values of $\zeta$ in these various cases.

Alongside the two distinctions already made (apolar versus polar, and contractile versus extensile), all rodlike molecules, including passive ones, fall into two further categories, known as ``flow aligning'' and ``flow tumbling''. Without activity, the former exhibit stable flow in which the director is inclined to the flow direction at a certain angle (the Leslie angle) whereas the latter undergo continuous director evolution, which is frequently chaotic~\cite{Chakrabarti04}.
To avoid the double complexity created by the flow tumbling instability on top of activity, we address here only the flow-aligning case.

Early theoretical work determined the linear stability of these systems
and it was found that an infinite sample of active material with nonzero $\mathbf{Q}$ (whether polar or apolar)
is hydrodynamically unstable to order parameter fluctuations, and that
this instability is connected to the generation of a spontaneous fluid 
flow~\cite{Hatwalne04,Kruse04,Simha02}. 
(For extensile -- but not contractile -- particles, this instability is present even if the flow velocity is constrained to be a function of one spatial coordinate only.)
It was also realised that the
introduction of boundaries together with suitable ``anchoring conditions'' (fixing the molecular orientation)
would lead to the stabilisation of the non-flowing ordered phase of uniform $\mathbf{Q}$. (Stability is restored for
small values of $\zeta$, lying below a threshold $\zeta_c$ which decreases
with system size, $L$, as $1/L^2$.) However, within these analytical approaches, it was not possible to determine the ultimate flow pattern resulting from this hydrodynamic instability. 

The simulations reported in \cite{Marenduzzo07a,Cates08,Marenduzzo07b}
gave therefore the first quantitative predictions for the resulting spontaneous flow patterns in unstable active fluids.
We focus first on a quasi-1D slab geometry with planar anchoring along
the boundary. Here it was found that for an apolar fluid,
upon increasing the activity, the system organises into a spontaneous 
Poiseuille flow (with a smoothly varying flow velocity, maximal at the centre of the slab). This spontaneously breaks symmetry and causes a net mass flux along the slab axis. For stronger activity and/or larger system sizes, one can also find spontaneously ``shear-banded'' flows in which successive layers of material have very different shear rates. 
 
In a 2D (thin film) flow geometry, the patterns differed significantly from the 1D case and were in good
qualitative agreement with observations of flow patterns in, for instance, 
concentrated {\it B. subtilus} and {\it Salmonella} 
suspensions~\cite{Harshey,Goldstein,Aranson,Cisneros07}. 
Fig. 2 shows some of the patterns which were found in 2D. In the apolar case, 
for values of $\zeta$ just above the threshold, the instability first evolves 
into steady state ``convective'' rolls (Figs. 1a-b)~\cite{Marenduzzo07b}. 
In some cases we also observed active bands or a succession of rolls (Fig. 2c),
whereas deeper in the active phase we found that an initial array of rolls 
breaks up into what looks chaotic flow (``bacterial turbulence'') at low Reynolds number (Fig. 2d-f). 
The equations of motion for polar active suspensions (see Appendix) can be treated with the same method.
This is somewhat similar, in that above a threshold there
is a transition to a spontaneously flowing state. In this case, however,
we do not observe rolls or bands and the system jumps directly into the
``turbulent'' flowing state (see Fig. 2f-h, with 2h showing trajectories
of tracer particles which highlight the chaotic appearance of the flow).

LB simulations have also proved helpful to characterise the rheology
of active fluids. In Ref.~\cite{Liverpool} it was suggested
that the viscosity of a contractile active fluid should diverge at
the passive isotropic-nematic transition in 2D. Simulations have
confirmed and generalised this to the case of a 3D order parameter (albeit constrained to undergo a 1D flow) where it was
clarified that the divergence there occurs at the spinodal point
(which unlike the 2D case is different from the 
isotropic-nematic transition point). LB simulations also showed that
upon increasing the density towards this spinodal point, one should observe a {\em decrease}
in the viscosity for extensile fluids.

The non-linear rheology should show even more striking behaviours~\cite{Cates08}.
Contractile fluids should strongly shear thicken for small forcing, 
and the extent of
thickening should depend on the distance from the isotropic-nematic transition. Within a bulk nematic phase, a formal yield stress (a stress threshold below which there is no flow in steady state) is predicted, whereas
for larger forcing, these materials exhibit shear thinning and approach the unenhanced (passive)
viscosity at very large shear rates. 

In contrast, isotropic suspensions of concentrated extensile particles (bacteria) should start from a low viscosity and thicken to
again approach the passive behaviour upon increasing the shear rate. Strikingly, a bulk oriented (nematic) phase of extensile active particles should show a zero effective viscosity for shear rates below some critical value.
This is because the 3D ordered system in the absence of stress (but constrained to have 1D flow) spontaneously organizes into two shear bands with flow in opposite directions, so that there is zero net velocity of the fluid. A finite relative velocity of the confining walls can now be accommodated, still at zero stress, by adjusting the relative amounts of the two bands. (Since there is now a finite shear rate and no shear stress, the viscosity is formally zero.) Finally, our studies suggested that when subjected to shear flow, 
extensile fluids should form shear bands more readily than passive ones close to the 
isotropic-nematic transition, while contractile fluids should shear-band less readily. 

These rheological predictions remain provisional, assuming as they do a 1D flow profile and the results may well be modified when this assumption is relaxed. (Such work is now underway in our group.) Nonetheless, the recent advances in experimental techniques, which have made it
possible to grow thin concentrated films of, e.g. {\it B. subtilis},
should render several of our results, such as the 2D predictions of Fig.2, testable in the near future.

\section{Blue phase simulations}

The ``blue phases'' (BPs) of chiral nematogenic molecules offer spectacular
examples of functional soft matter; each comprises a self-sustained network of 
disclinations~\cite{Wright89} embedded within a nematic matrix. (A disclination is a topological defect line, defined such that the nematic director rotates through a half-turn on treversing any circuit that encloses this line.) 

At high temperature, a fluid of chiral nematogenic molecules
remains in the ``isotropic phase'', with no preferred
orientation of the molecular axes.  Upon cooling down the sample, 
the molecules become oriented ($\mathbf{Q}$ is finite), but due to molecular chirality the director field $\mathbf{n}$ rotates with spatial position, describing everywhere a helix with a well-defined axis. (This is called the cholesteric phase.) However, very close to 
the transition, it is more 
advantageous locally for the director field to rotate in a helical fashion about any 
axis perpendicular to a straight line -- this complicated pattern was named 
a ``double twist cylinder'' \cite{Wright89}. Mathematically, it is impossible to patch together such double twist cylinders without creating defects in between, and this frustration 
gives rise to the disclination network observed in the blue phases, and 
responsible for many of their remarkable physical properties. Most striking among these are their optical properties (such phases can be made in all colours, not just blue) which stem from the presence of a lattice of disclinations with a unit cell whose size is comparable to the wavelength of visible light. In BP I and II (the two most common) this lattice has long range cubic order; BP III however is certainly not cubic, and probably not ordered. 

BPs have a fascinating scientific history. They were first reported in the 
late 19th century, by a scientist named Reinitzer, and then long forgotten, 
until some new experimental interest arose in the 1960s and 1970s.
Initial theories of BPs only came out in the 1980s (see e.g. \cite{Meiboom81}),
when the concept of double twist was first proposed. At this stage BPs were widely considered to be of purely academic interest, mainly because they were
only stable in a very narrow temperature range (about 1 K) close to the 
isotropic-cholesteric transition. In 1983, a world-leading expert on liquid 
crystals, F. C. Frank, said: 
``They [Blue phases] are totally useless, I think, except for one 
important intellectual use, that of providing tangible examples of topological 
oddities, and so helping to bring topology into the public domain of science, 
from being the private preserve of a few abstract mathematicians and particle
theorists.''~\cite{Wright89}.
In the first decade of the 21st century, this view rather suddenly changed, following advances in fabrication that enormously increased the stability range 
of BPs, up to about 50 K~\cite{Kikuchi02,Coles05}. In May 2008
Samsung presented the first blue-phase based liquid crystal display at 
the annual SID International Symposium. This new display is able to operate at a high frame frequency (240 Hertz), does not require costly alignment treatment at the boundaries of the liquid crystal, and may one day supersede current LC (twisted nematic) display technologies.

Traditional theories of BPs (see e.g.~\cite{Grebel84}) were based on 
semi-analytical approximations. While they were extremely useful to gain a 
qualitative understanding of the physics of BPs, these approaches had to rely 
on severe assumptions for progress to be possible. For instance, when 
estimating the phase diagram, the {\bf Q} tensor was approximated by a 
truncated Fourier series. Computational constraints only allowed very few
harmonics to be considered. As a result, when important experimental 
observations were mispredicted by the theory, it was not clear whether 
this was a drawback of the underlying free energy functional (the Landau-de Gennes free energy, see Appendix), or 
simply an artifact of the simplifications which were employed. These early 
theoretical works left unexplained the detailed shape of the phase 
diagram -- BPI and BPII appeared in the wrong order on increasing the
chirality $\kappa$. (This parameter is proportional to the inverse helical pitch of the cholesteric; see Appendix.) 
With an electric field applied, the analytic theories were also unable to account for 
the anomalous field-induced distortion (electrostriction) of BPI,  nor to explain why 
a new phase, named BPX, should be stable at all. Finally, most theories
assume cubic symmetry, so cannot describe BPIII, or the ``blue fog'', which
is thought to comprise a network of disclination lines without long-range order.

In recent years, LB simulations have been remarkably useful in filling most of these
conceptual theoretical gaps, and have significantly extended our quantitative 
understanding of the physics of blue phases. Firstly, in 
Refs.~\cite{Dupuis05a,Alexander08}, Eq. \ref{Q_motion} was solved by non-hybrid LB in the 
absence of fluid flow ($\mathbf{u}= \mathbf{0}$) with a set of initial conditions suggested from 
analytical expressions for the infinite chirality limit of BPI, BPII and 
$O_5$. ($O_5$ is another disclination lattice, which was proposed as stable by early 
theories, but not observed in experiments.) This procedure amounts to a free energy 
minimisation with a topological constraint; that is, the purely relaxational dynamics of Eq. \ref{Q_motion} was shown to maintain the point group of the disclination network chosen initially.
Therefore, this approach can be used to map out 
the full phase diagram as a function of chirality, $\kappa$, 
and  a reduced temperature parameter $\tau$ 
(see Appendix for the mathematical definition of these paramaters). 
This approach has the advantage of not making any approximations beyond those implicit in the selected (Landau-de Gennes) free energy. 

These LB simulations showed that the phase diagram predicted by the continuum
theory is actually in {\em good} qualitative agreement with the experiments:
BPI and BPII show up in the right order on increasing chirality, and $O_5$
is relegated to unphysical regions in the phase diagram~\cite{Dupuis05a}. The structure of these phases is shown in Fig.3a,b where a surface is drawn around each defect line at a certain contour of the ordering strength. (This creates a rendering of each disclination as a fattened tube.) The numerical phase diagram (Fig.3c) is in good agreement with the experimental one (Fig.3d).
An interesting calculation in \cite{Alexander06} has shown that even
including an extra set of spherical harmonic in the analytical scheme of
Ref. \cite{Grebel84} is unable to reproduce the details shown by the numerics.
This is an example in which simulations are extremely important to 
accurately find what the predictions of a given theory are. 

Similarly, LB simulations were performed in the presence of an 
electric field in Ref.~\cite{Alexander08}. It was found that under a small
electric field the unit cell of BPII tended to elongate along the field
direction, and shrink perpendicularly to it, whereas BPI displayed an opposite
behaviour, once more in agreement with experiments. An intermediate field
also turned the disclination network of BPI into a different structure (which it is tempting to identify with the experimentally observed BPX). Therefore, also when an
electric field is present, the Landau-de Gennes free energy works
remarkably well, although analytically tractable approximations of the resulting equations are not adequate to capture its predictions.

As well as allowing one to find equilibrium states under different fields and thermodynamic conditions, LB of course comes into its own for dynamical problems in which fluid flow cannot be ignored. For instance, the existence of a disclination network affects the 
response of a BP to an imposed Poiseuille flow~\cite{Dupuis05b}. Here,
LB work has shown that flow can lead to the unzipping of disclinations of 
integer topological charge. (On a circuit around such a defect line, the director rotates through a whole number of turns, rather than the half-turn around a standard disclination; these higher-order lines can form metastable 
networks in the absence of an applied flow.) Flow also can cause 
the bending and twisting of the BPI and BPII disclination networks. This bending
and twisting lead to an elastic component in the rheological behaviour, 
and as a result the simulations predict ``permeative'' flows, in which
the molecules comprising the blue phase flow through a static disclination pattern whose geometry is hardly perturbed by the underlying molecular transport.
The same kind of flow also occurs when cholesterics are sheared by
small forcing along
the direction of their helical axis~\cite{permeation}. Typically, BPs also 
display significantly shear thinning behaviour, as a strong enough flow
disrupts the disclination network in a manner that reduces the elastic stress.

Until very recently, LB work on BPs was limited to one unit cell of the
disclination lattice, within which 
several disclination cores are present and require a fine enough 
discretisation to be correctly resolved. However, supercomputers now allow
supra-unit cell simulations of BPs, the first account of which we have
recently given in Ref.~\cite{Henrich09a}, where we studied the domain
growth dynamics of a BPII domain inside a cholesteric or isotropic ``slab'',
in a parameter region in which the BP is the thermodynamically stable 
state. The simulations give evidence of an intriguing domain growth
kinetics. For small values of the chirality, the growth is slow and 
the resulting blue phase has no or few defects. When the chirality 
exceeds a certain threshold, however, the advancing disclination network
changes its symmetry and reconstruct into a new hexagonal phase, which is so
far undocumented in experiments. (This process is shown in Fig. 3e,f.)
It would be of interest to determine whether
this new BP is a metastable structure found due to the geometry we have
focussed on (we considered just one planar slice of unit cells) or
has a wider physical meaning. In all cases, these simulations are encouraging
as they suggest that large scale simulations of BPs are within reach. This
is of course of interest to the modelling of real devices, which
can be manufactured in the micron scale which we can consider computationally.

\section{Conclusions and future prospects}

We hope that this selection of results has shown that LB simulations
of liquid crystals are potentially extremely powerful in gaining new 
insights into the physics which is contained in the hydrodynamic equations
of motion of liquid crystalline fluids. 
For both active fluids and blue phases, it would have been
very difficult to compare theory and experiments -- even at a qualitative level -- without using these simulations. Although such comparisons remain in their infancy for active nematics, the spontaneous flow patterns in (for instance) concentrated bacterial suspensions, simulated via the 
continuum equations of motion of an active liquid crystal, are qualitatively comparable with the patterns observed in the experiments.
In the case of blue phases, such comparisons are more clear cut. Here it was thought that the classic theory based on a Landau-de Gennes
free energy was missing some physics because (for example) the phase
diagram was poorly predicted and anomalous electrostriction in BPI was
not found. Remarkably, LB has shown that this was a drawback of the
approximations used to make analytical progress, and not
of the original theory, which is qualitatively and semi-quantitatively accurate.

The fields we have covered in this short review are, of course, still full
of open questions, and we hope that future LB simulations will play an
important role in clarifying some of these. 

In active fluids, it will be important to characterise the flow patterns in fully three-dimensional active suspensions, and also to extend the treatment we have covered 
here to the case in which there are density fluctuations or inhomogeneity
in the fluid. Another related issue would be to study ``active emulsions'', 
in which droplets of active gels are suspended in an aqueous passive medium, 
possibly enclosed by an elastic membrane. Ultimately, it would be
very exciting if continuum theories like the one we solved numerically
may be applied to,
for instance, suspensions of cell extracts in an extracellular matrix.
From the theory point of view, it appears that an urgent issue is to
clarify to what extent active fluids faithfully represent concentrated
suspensions of motile particles, or swimmers, by, for instance, comparing the 
results of
continuum simulations to those of more microscopic models with fully
resolved swimmers, which can also be treated via LB (though of a different
kind than the one presented here)~\cite{Ishikawa08,Llopis06,Nash08}.

Large scale simulations of blue phases will also be likely to be important
in the future. From an application point of view, the exciting potential
of BP devices can ultimately be fully exploited if we manage to
reach a quantitative understanding of their thermodynamics, their 
switching dynamics, and the role of flow. Supra-unit cell
simulations are needed to this end, because the field leads to
unit cell deformations and may cause full scale reconstruction of the
disclination network. From a more fundamental point of view, we
do not have a satisfactory understanding of non-cubic blue phases.
Most notably, the structure of BPIII -- the ``blue fog'' -- 
is still not understood to date, and we hope that large scale simulations of 
amorphous disclination networks may shed some light on this elusive problem.

We are grateful to G. P. Alexander, S. M. Fielding, A. N. Morozov, 
E. Orlandini and J. M. Yeomans for useful discussions. 
We acknowledge EPSRC grants EP/E045316/1 and EP/E030173/1 for funding, 
and computer time on Hector funded by EP/F054750/1. MEC holds a Royal Society 
Research Professorship.

\clearpage
\begin{list}{}{\leftmargin 2cm \labelwidth 1.5cm \labelsep 0.5cm}

\item[\bf Fig. 1] (a) Turbulence in a sessile droplet of {\it B. subtilis}, viewed from
below a petri dish. The horizontal line is the edge of the droplet (picture
taken from Fig. 3 of Ref.~\cite{Goldstein}). The scale bar is 35 $\mu$m. 
(b) Flow pattern in a
similar bacterial droplet -- the arrow at the right stands for a speed of
35 $\mu$m/s (picture taken from Fig. 4 of Ref.~\cite{Goldstein}).

\item[\bf Fig. 2] Selected results from active fluid simulations. We only
plot the velocity field, resulting from LB solutions of
Eq.~\ref{Navier-Stokes}. 
The top row (a-c) shows stationary states obtained for 
apolar extensile fluids with moderate activity: it can
be seen that the spontaneous flow has the shape of rolls (a,b) or of
bands, in general tilted (c). The middle row show
non-stationary ``turbulent'' solutions for larger values of
the activity. 
The bottom row shows solution of the equations of motion of polar
active gels. In (g) there is no self-advection term, so that
the fluid is equivalent to an apolar gel, whereas in (h-i) this
term is switched on. In (i) we plot the trajectories of 3 tracer
particles, which show the ``turbulent'' nature of the flow.
Parameters
common to the apolar runs in (a)-(f) are: $\gamma=3$ (ensuring that we work
in the ordered phase), $\xi=0.7$ (which together with our
choice of $\gamma$ selects flow-aligning liquid crystals),
$\Gamma=0.33775$, $K=0.08$, $\eta=0.57$. The activity
parameter $\zeta$ was 0.001 (a), 0.002 (b), 0.01 (c), and
0.04 (d-f). Parameters for the polar runs in (g-i) are
$\zeta=0.02$, $\lambda=1.1$, $K=0.04$, $\Gamma=0.3$, $\eta=1.67$, and
$w=0$ (g), or $w=0.01$ (h,i).

\item[\bf Fig. 3] 
The top row shows the disclination lattices of BPI (a) and BPII (b),
as obtained from LB simulations -- 2 unit cells in each directions
are shown. The second row shows a computational (c, within the one
elastic constant approximation), and an experimental (d) typical
phase diagram for blue phases (picture taken from Fig. 3 of
Ref.~\cite{Crooker89}).
The key feature is that LB simulations
predict the correct order of appearance of BPI and BPII upon increasing
the chirality. Note that BPIII is not a cubic phase hence is not
included in the theoretical phase diagram. The bottom two rows (e-f)
show dynamical states obtained when a BPII domain grows inside
an initially cholesteric matrix (see Ref.~\cite{Henrich09a}).
The reduced temperature was $\tau=0$ and the chirality was 
$\kappa=2$ (e) or $\kappa=1$ (f).
\end{list}

\clearpage

\begin{figure}[ht]
\begin{center}
\includegraphics[width=1.0\textwidth]{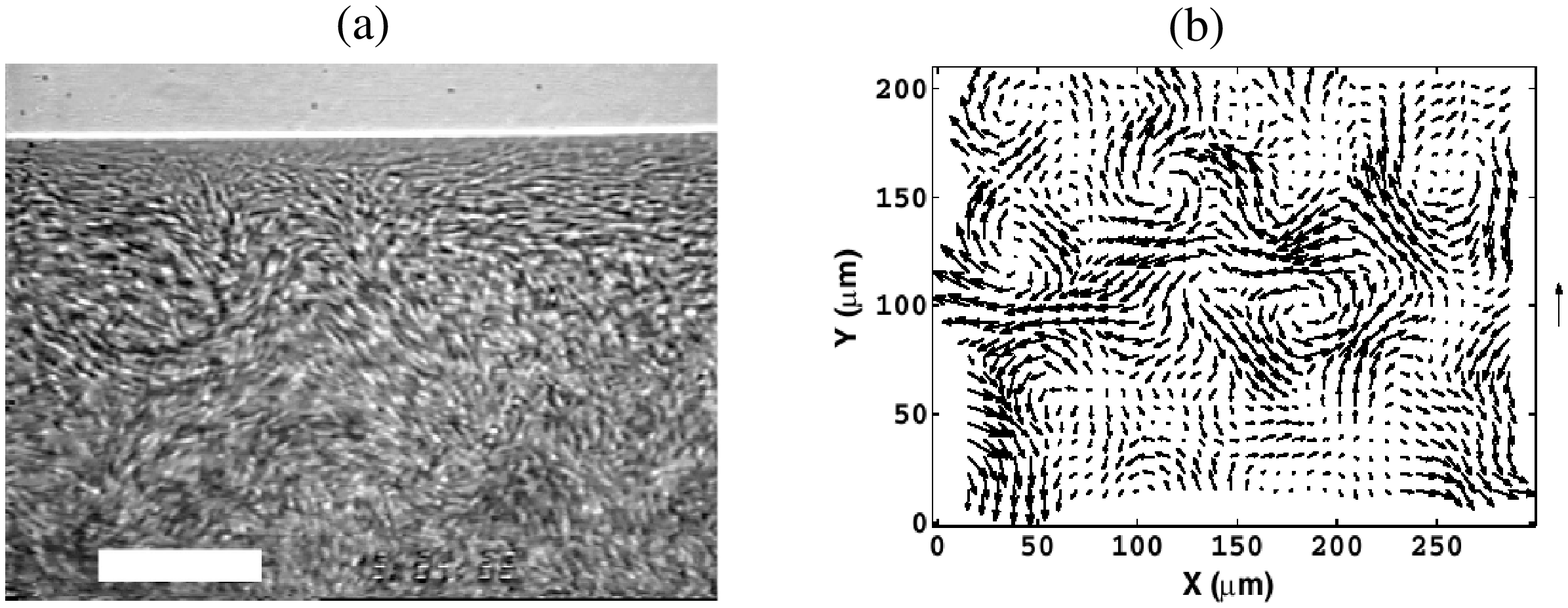}
\caption{}
\end{center}
\end{figure}

\clearpage

\begin{figure}[ht]
\begin{center}
\includegraphics[width=1.0\textwidth]{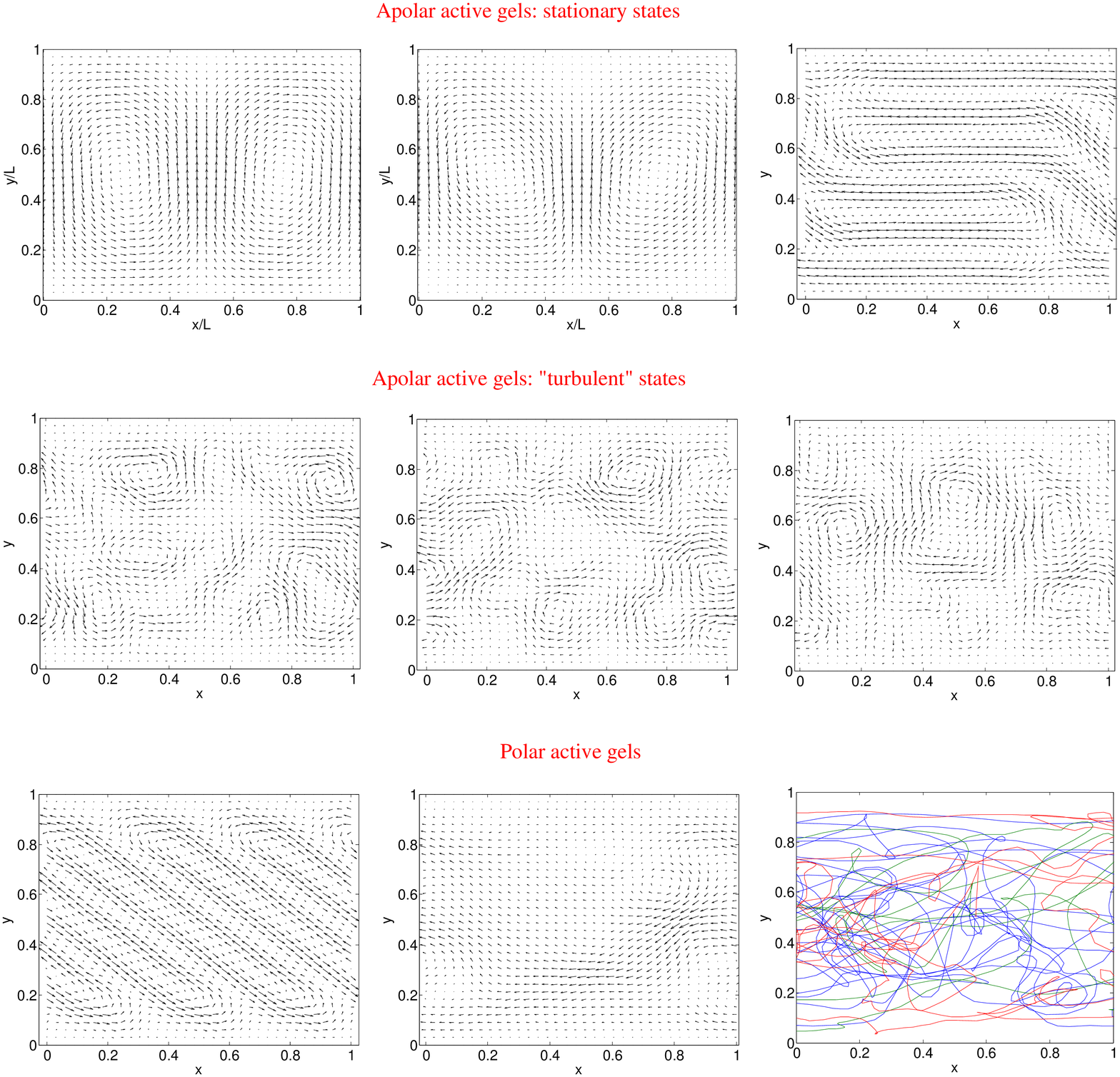}
\caption{}
\end{center}
\end{figure}

\clearpage

\begin{figure}[ht]
\begin{center}
\includegraphics[width=0.7\textwidth]{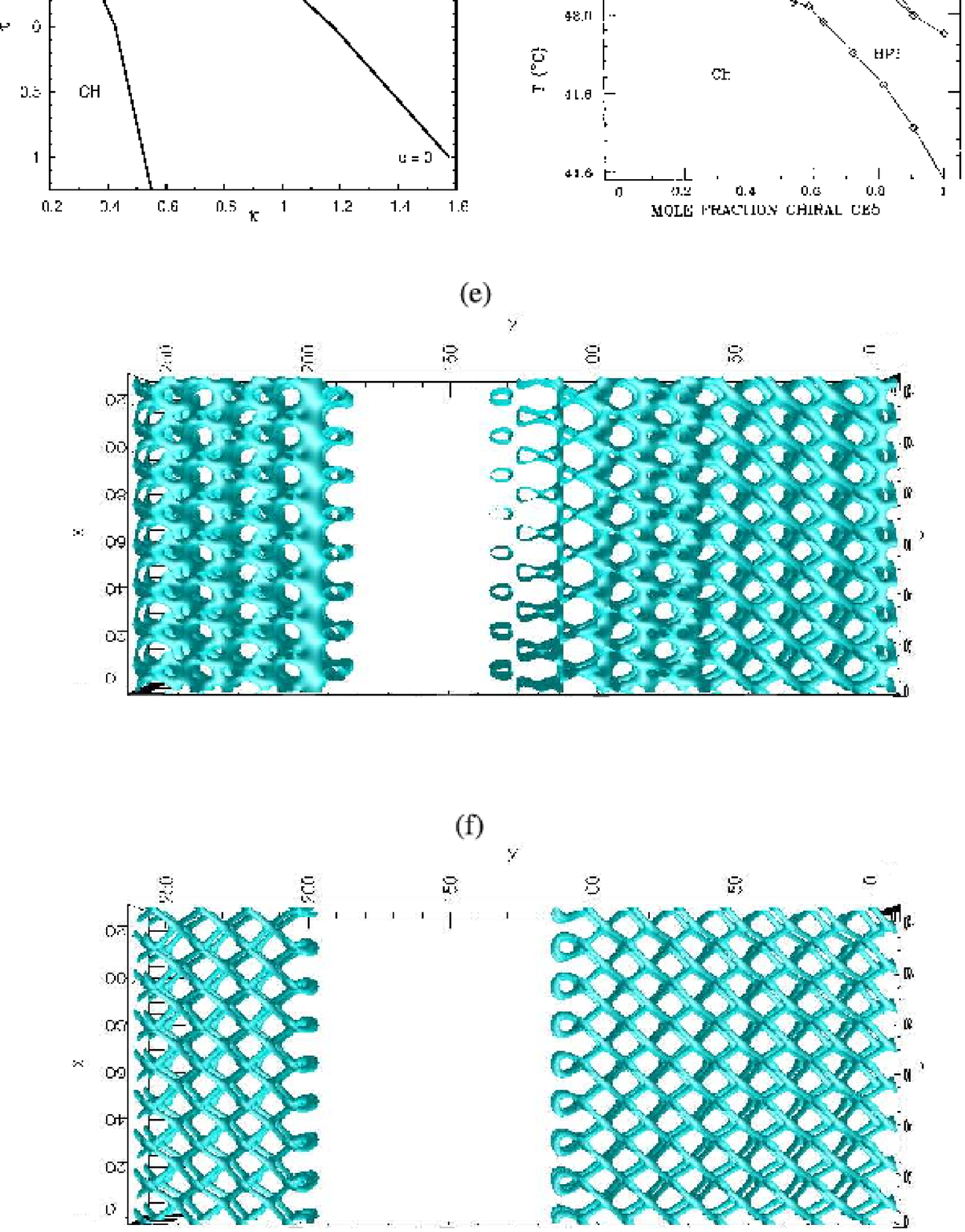}
\caption{}
\end{center}
\end{figure}

\clearpage

\appendix

{\bf Appendix: Hydrodynamic equations of motion for active and passive liquid
crystalline fluids}\\

In this Appendix we review the equations of motion for (active and passive)
liquid crystalline fluids, which we solve by lattice Boltzmann simulations. 
These are the equations used to generate the results reviewed in our work.

We first describe the thermodynamics of a liquid crystalline fluid in the absence of active stresses. This covers 
cholesterics and blue phases (and also active gels within a passive phase). 
We employ a Landau-de Gennes free energy ${\cal F}$, whose density
we indicate by $f$. The free energy density can be written as a sum of
two contributions, $f_1$ and $f_2$. The first is a bulk contribution,
\begin{eqnarray}\nonumber
f_1 & = & \frac{A_0}{2}(1 - \frac {\gamma} {3}) Q_{\alpha \beta}^2 -
          \frac {A_0 \gamma}{3} Q_{\alpha \beta}Q_{\beta
          \gamma}Q_{\gamma \alpha} \\ \nonumber
& + & \frac {A_0 \gamma}{4} (Q_{\alpha \beta}^2)^2,
\label{eqBulkFree}
\end{eqnarray}
while the second is a distortion term.
For nonchiral liquid crystals, we take the 
(standard) one elastic constant approximation~\cite{Beris}

\[
f_2=\frac{K}{2} \left(\partial_\gamma Q_{\alpha \beta}\right)^2.
\]

Where $A_0$ is a constant, $\gamma$ controls the
magnitude of order (it may be viewed as an effective
temperature or concentration for thermotropic and lyotropic
liquid crystals respectively), while
$K$ is an elastic constant.
To describe cholesterics, we employ the slightly generalised distortion
free energy, which is again standard~\cite{Wright89}:
\[
f_2=\frac{K}{2} 
\left[ \bigl(\partial_{\beta}Q_{\alpha\beta}\bigr)^2+
           \bigl( \epsilon_{\alpha \gamma \delta} \partial_{\gamma} Q_{\delta \beta} 
           + 2q_0 Q_{\alpha \beta} \bigr)^2\right].
\]
Here and in what follows Greek indices
denote cartesian components and summation over repeated indices is implied.

For blue phases, it is customary to identify the position in thermodynamic parameter
space via the chirality, $\kappa$, and the
reduced temperature, $\tau$. These may be defined in terms of previous
quantities via~\cite{Alexander06,Alexander08}:
\begin{eqnarray}\nonumber
\kappa & = & \sqrt{\frac{108 K q_0^2}{A_0\gamma}} \\ \nonumber
\tau & = & 27 \left(\frac{1-\gamma/3}{\gamma}\right).
\end{eqnarray}
(Note that the reduced temperature was defined in older literature
as $\tau = 27 \frac{1-\gamma/3}{\gamma}+\kappa^2$.) 

When needed, the anchoring of the director field on the boundary surfaces
(Fig. 2) to a chosen unit vector ${\bf n}^0$
is ensured by adding a surface term in the free energy density
\begin{eqnarray}\nonumber
f_s & = & \frac{1}{2}W_0 (Q_{\alpha \beta}-Q_{\alpha \beta}^0)^2\\
\nonumber
Q_{\alpha \beta}^0 & = & S_0 (n_{\alpha}^0n_{\beta}^0-\delta_{\alpha\beta}/3)
\end{eqnarray}
The parameter $W_0$ controls the strength of the anchoring, while $S_0$
determines the degree of the surface order. If the surface order is to equal
the bulk order, $S_0$ should be set equal to $q$, the order
parameter in the bulk ($3/2$ times the largest eigenvalue of the {\bf Q}
tensor). $W_0$ is large (strong anchoring) in what follows.

The equation of motion for {\bf Q} is taken to be \cite{Beris}
\[
(\partial_t+{\vec u}\cdot{\bf \nabla}){\bf Q}-{\bf S}({\bf W},{\bf
  Q})= \Gamma {\bf H}+\tilde\lambda {\bf Q}
\]
where $\Gamma$ is a collective rotational diffusion constant,
and $\tilde\lambda$ is an activity parameter which for simplicity we set to zero in our simulations. (The resulting term can anyway be absorbed into a shift of $A_0$ and/or $\gamma$ in the free energy.) 
The first term on the left-hand side of the equation above
is the material derivative describing the usual time dependence of a
scalar quantity advected by a fluid with velocity ${\vec u}$. This is
modified for rod-like molecules by a second term
\begin{eqnarray}
\nonumber
{\bf S}({\bf W},{\bf Q})
& = &(\xi{\bf D}+{\bf \omega})({\bf Q}+{\bf I}/3) \\ \nonumber
& + & ({\bf Q}+
{\bf I}/3)(\xi{\bf D}-{\bf \omega}) \\ \nonumber
& - & 2\xi({\bf Q}+{\bf I}/3){\mbox{Tr}}({\bf Q}{\bf W})
\end{eqnarray}
where Tr denotes the tensorial trace, while
${\bf D}=({\bf W}+{\bf W}^T)/2$ and
${\bf \omega}=({\bf W}-{\bf W}^T)/2$
are the symmetric part and the anti-symmetric part respectively of the
velocity gradient tensor $W_{\alpha\beta}=\partial_\beta u_\alpha$.
The constant $\xi$ depends on the molecular
details of a given liquid crystal, and determines, together with
$\gamma$, whether a liquid crystal is flow aligning or flow tumbling
(we restrict to the former case in this work).
The first term on the right-hand side of the order parameter evolution
equation describes the relaxation of the order parameter towards the minimum of
the free energy. The molecular field ${\bf H}$ which provides the driving
motion is given by
\[
{\bf H}= -{\delta {\cal F} \over \delta {\bf Q}}+({\bf
    I}/3) {\mbox{Tr}}{\delta {\cal F} \over \delta {\bf Q}}.
\label{molecularfield}
\]

The fluid velocity, $\vec u$, obeys the continuity equation and
the Navier-Stokes equation, whose incompressible limit is Eq. \ref{Navier-Stokes}
\if{
\[
\rho(\partial_t+ u_\beta \partial_\beta)
u_\alpha = \partial_\beta (\Pi_{\alpha\beta})+
\eta \partial_\beta(\partial_\alpha
u_\beta + \partial_\beta u_\alpha)
\]
}\fi
in which $\Pi_{\alpha\beta}=\Pi^{\rm passive}_{\alpha\beta}+
\Pi^{\rm active}_{\alpha\beta}$. The stress tensor
$\Pi^{\rm passive}_{\alpha\beta}$ necessary to describe ordinary LC
hydrodynamics is (up to an isotropic pressure term) given by:
\begin{eqnarray}\nonumber
\Pi^{\rm passive}_{\alpha\beta}& = & 2\xi
(Q_{\alpha\beta}+{1\over 3}\delta_{\alpha\beta})Q_{\gamma\epsilon}
H_{\gamma\epsilon}\\\nonumber
&-&\xi H_{\alpha\gamma}(Q_{\gamma\beta}+{1\over
  3}\delta_{\gamma\beta})\\ \nonumber
&-&\xi (Q_{\alpha\gamma}+{1\over
  3}\delta_{\alpha\gamma})H_{\gamma\beta}\\ \nonumber
&-&\partial_\alpha Q_{\gamma\nu} {\delta
{\cal F}\over \delta\partial_\beta Q_{\gamma\nu}}
\\ \nonumber
& + & Q_{\alpha \gamma} H_{\gamma \beta} -H_{\alpha
 \gamma}Q_{\gamma \beta}.
\label{BEstress}
\end{eqnarray}
whereas the active term is given by, in leading order
\[
\Pi^{\rm active}_{\alpha\beta}=-\zeta  Q_{\alpha\beta}
\]
where $\zeta$ is an activity constant \cite{Hatwalne04}.
Note that with the sign convention chosen here $\zeta>0$  
corresponds to extensile rods and $\zeta<0$ to
contractile ones \cite{Hatwalne04}.

\if{
A full understanding of the microscopic origin of the
phenomenological couplings $\zeta$ and $\lambda$, as well
as of the range of values these may attain in physically relevant
situations, will require multi-scale modelling at different coarse 
graining levels, and more accurate quantitative experiments.
These are at the moment still lacking. However, we 
already know from experiments and from some
more microscopic approaches, that actomyosin gels are
contractile, so that in physiological conditions those materials
should be described by negative values of $\zeta$ \cite{Liverpool}. 
Bacterial solutions, on the other hand, are typically
extensile~\cite{Hatwalne04,Ishikawa08}, although exceptions exist. The 
term proportional to $\lambda$ has been proposed in Ref. 
\cite{Hatwalne04} as a symmetry allowed term which, for
dilute bacterial suspensions, should be negative and proportional to
the inverse of the time scale for relaxation of 
activity-induced ordering. 
}\fi


We can also use a variant of these equations to study polar active gels. The order parameter is this time a vector $P_\alpha$ (with variable magnitude) as there is no longer head-tail symmetry in
the system. The equations of motion we used in our LB simulations
(reported in Fig. 2, bottom row), are a simplified version of those
presented in \cite{Giomi08}.
The equation governing the evolution of the vectorial order parameter is 
\begin{eqnarray}\nonumber
\left[\partial_t+\left(u_{\beta} +wP_{\beta}\right)\partial_{\beta})\right]
P_{\alpha}= \\ \nonumber 
\lambda D_{\alpha\beta} P_{\beta} -\omega_{\alpha\beta}P_{\beta}
+\Gamma' h_{\alpha}.
\end{eqnarray}
In this equation, $w$ is another active term, due to swimming, which causes
self-advection of the order parameter, while $\lambda$ is a material
dependent constant -- positive for rod-like molecules. If $|\lambda|>1$
the liquid crystalline passive phase is flow-aligning, otherwise it is
flow-tumbling. 
The ``molecular field'' is now given by 
$h_{\alpha}=-\delta {\cal F}_{\rm pol}/\delta p_{\alpha}$
where ${\cal F}_{\rm pol}$ is the free energy for a polar active nematic,
whose density is (see also Ref.~\cite{Giomi08} where a more
general form is used):
\begin{equation}
f =\frac{a}{2} |{\mathbf P}|^2+\frac{b}{4}|{\mathbf P}|^4+
K\left(\partial_\alpha P_{\beta}\right)^2
\end{equation}
where $K$ is an elastic constant, $b>0$, and we chose $a=-b$ to ensure that
the minimum of the free energy is with $|{\mathbf P}|=1$. (Note that in
this model ${\mathbf P}$ this time denotes a vector rather than a tensor).

The Navier-Stokes equation is as in the tensorial model, but the
stress tensor this time is 
\[
\frac{1}{2}\left(P_{\alpha}h_{\beta}-
P_{\beta}h_{\alpha}\right)-\frac{\lambda}{2}\left(P_{\alpha}h_{\beta}+
P_{\beta}h_{\alpha}\right) 
- \zeta P_{\alpha}P_{\beta}.
\]
As mentioned in the text, the active term in the stress tensor, proportional
to $\zeta$, has therefore the same form for apolar and polar active gels. 

A few limits of the two theories considered above are worth noting.
The tensorial model we have written down is equal to, for $\zeta=0$,
the Beris-Edwards model for liquid crystal hydrodynamics. Analogously,
for $\zeta=w=0$ the polar model reduces to the Leslie-Ericksen model
of nematodynamics. For $w=0$, and a sample of uniaxial active 
liquid crystals, with a spatially uniform
degree of orientational order, the tensorial model may be mapped onto
the vectorial one (see \cite{Marenduzzo07b} for a proof of this).

\end{document}